\documentclass[aps,twocolumn,superscriptaddress,showkeys]{revtex4}
\usepackage{amsmath}
\usepackage{color,xcolor}
\usepackage{graphicx}
\newcommand{\bastar}{\begin{eqnarray*}}
\newcommand{\eastar}{\end{eqnarray*}}
\newskip\humongous \humongous=0pt plus 1000pt minus 1000pt

\newif\ifdtup

\relax
\newcommand{\W}{{\vec W}}
\newcommand{\n}{\hat n}
\newcommand{\hn}{\hat n}
\newcommand{\hr}{\hat r}

\newcommand{\hD}{{\hat D}}

\newcommand{\bea}{\begin{eqnarray}}
\newcommand{\eea}{\end{eqnarray}}
\newcommand{\pd}{\partial}

\newcommand{\Int}{\displaystyle\int}

\newcommand{\hB}{{\hat B}}

\newcommand{\tB}{{\tilde B}}
\newcommand{\tC}{{\tilde C}}

\newcommand{\G}{{\vec G}}
\newcommand{\B}{{\vec B}}

\newcommand{\hG}{{\hat G}}

\newcommand{\mn}{{\mu\nu}}

\newcommand{\eps}{\epsilon}
\newcommand{\beps}{\bar \epsilon}

\newcommand{\valpha}{{\vec \alpha}}

\newcommand{\lam}{\lambda}
\newcommand{\vsig}{{\vec \sigma}}
\newcommand{\nn}{\nonumber}
\newcommand{\pro}{\partial}

\begin{document}
\title{Abelian and Non-Abelian Monopole Configuration
in Condensed Matters}

\author{Yunkyu Bang}
\email{ykbang@apctp.org}
\affiliation{Asia Pacific Center for Theoretical Physics,
Pohang 37673, Korea}
\affiliation{Department of Physics, Pohang Institute
of Science and Technology, Pohang 37673, Korea}
\author{Y. M. Cho}
\email{ymcho0416@gmail.com}
\affiliation{Asia Pacific Center for Theoretical Physics,
Pohang 37673, Korea}
\affiliation{School of Physics and Astronomy,
Seoul National University, Seoul 08826, Korea}
\affiliation{Center for Quantum Spacetime,
Sogang University, Seoul 04107, Korea}
\author{Tieyan Si}
\email{tieyansi@hit.edu.cn}
\affiliation{Department of Physics, Harbin Institute
of Technology, Harbin 150080, China}
\author{Li-Ping Zou}
\email{zoulp@impcas.ac.cn}
\affiliation{Institute of Modern Physics, Chinese Academy
of Science, Lanzhou 730000, China}
\affiliation{School of Physics and Astronomy,
Sun Yat-Sen University, Zhuhai 519082, China}

\begin{abstract}
We discuss the Abelian and non-Abelian monopoles which 
could exist in condensed matters. We show how the Dirac 
monopole can be regularized by the charge screening, 
and argue that the Dirac monopole of mass of hundred 
meVs could exist in dielectric condensed matters. Moreover, 
we generalize this result to non-Abelian condensed matters
to show the existence of the non-Abelian monopole 
configuration in two-gap condensed matters, and present 
explicit monopole solutions. 
\end{abstract}
\pacs{11.15.Ex, 14.80.Hv, 74.25.Ha}
\keywords{topological objects in condensed matters,
regularization of Dirac monopole in condensed matters,  
non-Abelian monopole in two-gap condensed matters}
%%%%%%%%%%%%%%%%%%%%%%%%%%%%%%%%%
\maketitle

\section{Introduction}

Topological objects have played important roles in
physics. Late nineteen century Kelvin first suggested
that the atoms could be viewed as knots whose
stability could be explained by the topology \cite{kelvin}.
Later Dirac introduced the Dirac monopole based on
the non-trivial $U(1)$ topology \cite{dirac}. With this
topology and topological objects have played fundamental
roles in physics. In particular, the monopole has
become an obsession in theoretical and experimental
physics \cite{wu,thooft,julia,prl80,dokos,cab,medal,atlas,icecube}.

After the Dirac monopole, we have had the Wu-Yang
monopole, the 'tHooft-Polyakov monopole, and grand
unification monopole \cite{wu,thooft,julia,prl80,dokos}.
But the most realistic monopole which could exist
in nature is the electroweak monopole in the standard
model which can be viewed as a hybrid between
the 'tHooft-Polyakov monopole and Dirac
monopole \cite{plb97,yang,epjc15}. The importance
of this monopole comes from the following facts. First,
it appears as the electroweak generalization of Dirac
monopole. So it is this monopole, not the Dirac monopole,
which should exist in nature. Second, unlike the Dirac
monopole which is optional, the electroweak monopole 
must exist if the standard model is correct. This means 
that the discovery of this monopole, not the Higgs particle, 
should be viewed as the final test of the standard model.

Moreover, it has deep implications in cosmology. As
the only heavy and stable particle in the early universe
it could generate the primordial black holes which
could account for the dark matter, and become
the seeds of the large scale structures of the universe.
Most importantly, when discovered, it will become
the first magnetically charged and stable topological
elementary particle in the history of physics. For this
reason MoEDAL and ATLAS at LHC, IceCube at south pole, 
and similar experiments are actively searching for
the monopole \cite{medal,atlas,icecube}.

Topological objects have also played important roles in
condensed matter physics. The best known topological
object in condensed matter is the Abrikosov vortex in superconductor made of quantized magnetic flux, which 
comes from the $\pi_1(S^1)$ topology of the Abelian $U(1)$ 
gauge theory \cite{abri}. Similar string-like topological objects 
could also exist in non-Abelian two-component Bose-Einstein condensates and $^3He$ superfluids \cite{prb05,pra05}. 
Moreover, recently it has been shown that two-gap 
superconductors could have non-Abrikosov type magnetic 
vortices, D-type or N-type, with integer or fractional magnetic 
flux \cite{epjb08}.

While these are certainly interesting topological objects,
a more interesting object should be the monopole which 
has the $\pi_2(S^2)$ topology. There have been serious
efforts to search for monopole-like objects in the condensed 
matter systems. One is to identify the monopole as emergent excitations of point charges of dipole moment in spin
ice \cite{cmm1}. Another is to identify the monopole as
point-like topological defects in spinor Bose-Einstein
condensates \cite{cmm2}. A third approach is to treat
angulon quasi-particles of rotating molecules in superfluid
helium as an effective non-Abelian monopole \cite{cmm3}.

However, these monopoles are collective phenomena 
exhibiting the $\pi_2(S^2)$ topology, which may not be 
viewed as a genuine magnetic monopole. {\it The purpose 
of this paper is to argue that a genuine magnetic monopole 
could exist in dielectric and/or two-gap condensed matters, 
at least theoretically. We first show that a regularized Dirac monopole of the mass hundreds meV could exist in ordinary condensed matters. Next, we discuss under what condition 
a monopole-like configuration can exist in non-Abelian 
condensed matters, and construct explicit monopole solutions 
which could exist in a realistic two-gap condensed matter.}

In specific, we show how the charge screening in dielectric condensed matters can be applied to make the energy of 
Dirac monopole finite. Moreover, we generalize the Dirac 
monopole to non-Abelian monopole and present explicit 
solutions of monopole and dyon which carry a Dirac-type 
singularity in two-gap condensed matters, and show that this singularity can be regularized by the same mechanism, 
by a non-vacuum electric permittivity that mimics the quantum correction of charge renormalization by virtual electron-positron 
pair production. 

To generalize the Dirac monopole to a non-Abelian monopole,
we need to understand the relation between the Abelian
and non-Abelian gauge theories. A best way to understand  
this is the Abelian decomposition of non-Abelian gauge 
theory \cite{prd80,prl81}. It decomposes the non-Abelian 
gauge potential into the restricted Abelian part and 
the non-Abelian valence part gauge independently. Moreover, 
it shows that the restricted Abelian gauge potential is made 
of two parts, the naive Abelian (electric) potential and 
the topological monopole (magnetic) potential.

More importantly, the restricted potential retains the full 
non-Abelian gaugedegrees of freedom and thus inherits 
the full non-Abelian topology of the theory, in spite of 
the fact that it describes the Abelian sub-dynamics of 
the non-Abelian gauge theory. This means that we can 
construct the restricted gauge theory with 
the restricted gauge potential which has the full
non-Abelian gauge symmetry but describes the simpler
Abelian sub-dynamics of the non-Abelian gauge
theory \cite{prd80,prl81}.

In comparison, the non-Abelian valence potential
transforms covariantly under the gauge transformation 
and thus play the role of a gauge covariant source of 
the restricted potential. Thus the Abelian decomposition 
teaches us how the Abelian structure is embedded in 
the non-Abelian gauge theory and shows how to separates 
the Abelian structure from the non-Abelian gauge theory 
gauge independently. Moreover, it allows us to express 
the non-Abelian gauge theory in an Abelian form which is 
very useful for the condensed matter physics, as we will 
show in the following. 

The Abelian decomposition has played a crucial role 
for us to clarify the complicated non-Abelian dynamics 
in QCD \cite{plb82,prd00,jhep05,prd13,epjc19}. It 
decomposes the gluon to the color neutral neuron and 
colored chromon, and separates the topological monopole 
potential gauge independently. Moreover, it allows us 
to prove the monopole condensation necessary for the color confinement in QCD.

In this paper, we show that the Abelian decomposition
can also play important role in condensed matter physics. 
This is because it separates the topological structure
of the non-Abelian gauge theory gauge independently,
and allows us to construct the topological objects more 
easily. Moreover, it shows how the Abelian gauge
theory is related to the non-Abelian gauge theory by
``abelianizing" the non-Abelian gauge theory gauge
independently. This is very important, because condensed 
matters are often described by QED which is Abelian.

The paper is organised as follows. In Section II, we show
how the Dirac monopole can be regularized in ordinary 
condensed matters by the non-trivial electromagnetic 
permittivity, and argue that the monopole with mass 
of hundreds meV could exist in ordinary condensed matters. 
In Section III we discuss a general framework which 
describes non-Abelian condensed matters and its 
mathematical structure, in particular, its topological structure, 
using the Abelian decomposition. In Section IV, we show 
how the singular monopole and dyon could exist in realistic 
two-gap condensed matters and present explicit solutions. 
In Section V, we show how to regularize the singular 
monopole and dyon solutions with the real electromagnetic permittivity which describes the electric charge screening,
and show that in the Abelian limit the solution reduces to 
the regularized Dirac monopole. In Section VI, we compare 
our solutions with the monopole and dyon existing in 
the standard model. Finally in the last section we discuss 
the physical implications of our result.

\section{Regularized Dirac Monopole in Ordinary Condensed Matters}

The topological objects in condensed matters can be of
Abelian or non-Abelian. The best known topological object 
in Abelian gauge theory is the Dirac monopole. It is well 
known that the U(1) gauge theory which describes 
the Maxwell's theory has no magnetic monopole. 
This is because the Maxwell's equation, in particular 
the Bianchi identity, forbids the existence of the monopole. 
But in 1931 Dirac predicted the existence of the monopole generalizing the Maxwell's equation and making the U(1) 
gauge theory non-trivial \cite{dirac}. He showed that 
the Maxwell's theory can be generalized top admit 
the monopole, if we impose the charge quantization condition $eg=2\pi n$. Since there is no explanation why the electric 
charges in nature are quantized, this charge quantization 
rule has often been used to argue the existence of 
the monopole. Later Schwinger generalized the monopole 
to dyon \cite{schw}.

In spite of the huge efforts to try to discover the monopole experimentally, however, the monopole has not been 
discovered yet \cite{cab}. This was (at least) partly because 
there has been no information about the mass of 
the monopole. Since the Dirac monopole has infinite energy, 
there was no way to predict the mass. This has made 
most of the monopole experiments a blind search in the dark 
room, with no theoretical lead. 

Now we show the existence of the regularized Dirac 
monopole which might have a finite mass of the order 
of meV in dielectric condensed matters. Consider 
the following Maxwell's theory coupled to the neutral 
scalar field $\rho$,
\begin{gather}
{\cal L} =-\frac12 (\pd_\mu \rho)^2
-\frac{\lambda}{2}\big(\rho^2-\rho_0^2 \big)^2
-\frac{1}{4} \eps(\rho)~F_\mn^2,
\label{dlag}
\end{gather}
where we require $\eps$ approaches to one asymptotically
to make sure that the theory reduces to the Maxwell's theory. 
We can interpret $\rho$ as an emergent scalar field which represents the density of the polarization in condensed 
matters responsible for the charge screening. This is because 
effectively $\epsilon$ in front of $F_\mn$ (when it becomes 
a constant) changes the gauge coupling $e$ to the ``running" coupling $\bar e=e /\sqrt{\epsilon}$, so that with the rescaling 
of $A_\mu$ to $A_\mu/e$, $e$ changes to $e /\sqrt{\epsilon}$. 
This makes (\ref{lag0}) an ideal Lagrangian 
to describe the Dirac monopole in condensed matter.

To find the desired monopole solution choose the following 
ansatz for the Dirac monopole
\begin{gather}
\rho=\rho(r),   \nn\\
A_\mu= -\frac1e (1-\cos\theta)\pd_\mu \varphi.
\label{dmans}
\end{gather}
The ansatz has the well known Dirac string along 
the negative $z$-axis, but it can be gauged away and 
made unphysical when we make the U(1) bundle 
non-trivial and impose the charge quantization condition.

\begin{figure}
\includegraphics[height=4.5cm, width=8cm]{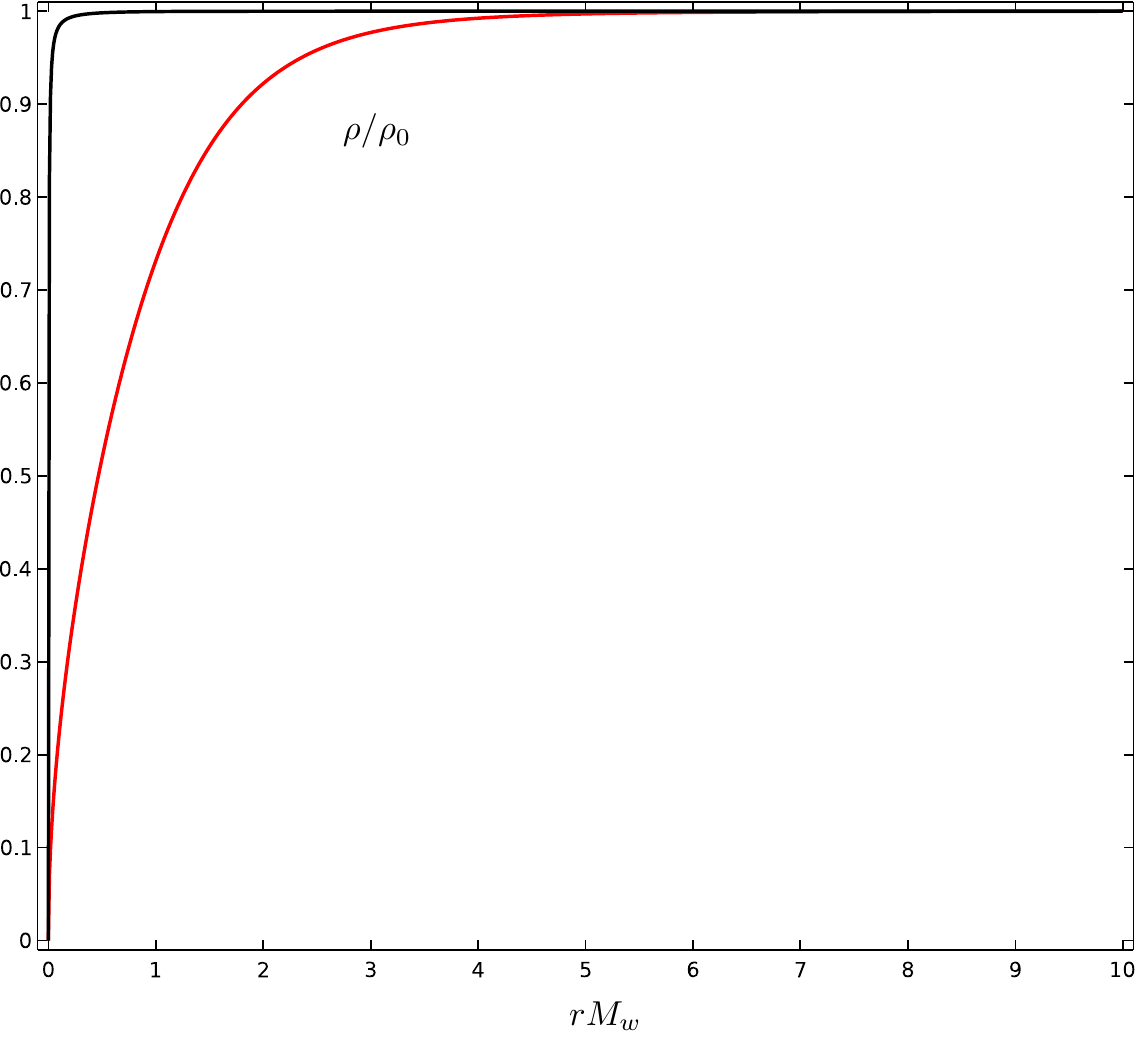}
\caption{\label{femh} The regularized Dirac monopole 
solution (the red curves) regularized by the real electromagnetic
permittivity $\eps=(\rho/\rho_0)^6$. Exactly the same solution 
also describes the finite energy monopole in two-gap 
condensed matters.}
\end{figure}

With this we have the equation of motion
\begin{gather}
\ddot \rho + \frac{2}{r}\dot \rho
=\frac{\lambda}{2} (\rho^2- \rho_0^2) \rho
+\frac{\eps'}{2} \frac{1}{e^2 r^4},
\label{mdeq}
\end{gather}
where $\eps' = d\eps/d\rho$. We can easily solve 
this with the boundary condition
\begin{gather}
\rho(0)=0,~~~~~\rho(\infty)=\rho_0, 
\end{gather}
and obtain the finite energy monopole solution with
\begin{gather}
\eps \simeq \Big(\frac{\rho}{\rho_0}\Big)^n.
\end{gather}
The solution for $n=6$ is shown in Fig. \ref{femh}. This 
tells that we can regularize the Dirac monopole, replacing 
the vacuum electromagnetic permittivity with a real 
electromagnetic permittivity. This is unexpected. As far as 
we understand, there has been no known way to regularize 
the Dirac monopole other than this. Exactly the same 
mechanism (the charge renormalization by the vacuum 
polarization) has been shown to make the mass of 
the electroweak monopole finite \cite{epjc19}. 

The monopole energy is given by 
\begin{gather}
E=4\pi \int_0^\infty dr \bigg\{\frac{\beps}{2e^2 r^2} 
+\frac12 (r\dot \rho)^2   \nn\\
+\frac{\lambda r^2}{8}\big(\rho^2-\rho_0^2 \big)^2 \bigg\},
\label{dde}
\end{gather}
which, in the limit $\lambda$ goes to zero, becomes
\begin{gather}
E \simeq 0.25 \times \frac{4\pi}{e}~\rho_0,
\label{dde1}
\end{gather} 
where $\rho_0$ is in principle arbitrary. This has deep 
implication. It is natural to assume that  $\rho_0$ in 
condensed matters to be of the order of the coherence 
length in ordinary superconductor, of the order of meV. 
In this case the Dirac monopole in ordinary superconductors 
could have mass about $1/\alpha$ times the coherence 
length, where $\alpha$ is the fine structure constant. 
This is because the monopole mass is given by 
$4\pi/e \rho_0$, while the coherence length is given by 
$e\rho_0$. This tells that the mass of the Dirac monopole 
in condensed matter could be of the order of hundred meV. 
This is really remarkable. In the following we will show 
that exactly the same regularized monopole could exist 
in two-gap condensed matters. 

\section{Non-Abelian Structure in Condensed Matters: General Framework}

Now, we discuss the non-Abelian monopole in condensed 
matters. Consider a complex doublet $\phi=(\phi_1,\phi_2)$ 
made of two condensates $\phi_1$ and $\phi_2$ which 
could represent two types of condensates similar to 
two different Cooper-pairs in two-gap superconductors 
or two types of states in two-component Bose-Einstein 
condensates \cite{pra05,prb05,epjb08}. The doublet could 
naturally accommodate the U(2) or SU(2)xU(1) gauge 
interaction described by
\begin{gather}
{\cal L} =-|{\cal D}_\mu \phi|^2
-\frac{\lambda}{2}\big(|\phi|^2
-\frac{\mu^2}{\lambda}\big)^2
-\frac14 F_\mn^2-\frac14 \G_\mn^2, \nn \\
{\cal D}_\mu \phi =\big(\pd_\mu-i\frac{g}{2} A_\mu
-i\frac{g'}{2} \vsig \cdot \B_\mu \big) \phi
=(D_\mu -i\frac{g}{2} A_\mu) \phi, \nn\\
D_\mu \phi=(\pd_\mu
-i\frac{g'}{2} \vsig \cdot \B_\mu \big) \phi,
\label{lag0}
\end{gather}
where $A_\mu$ and $\B_\mu$ are the overall U(1)
and SU(2) gauge potentials, $F_\mn$ and $\G_\mn$
are the corresponding field strengths, $g$ and $g'$
are the coupling constants.

One might wonder what is the motivation to consider
the above Lagrangian. Consider first the overall U(1) 
gauge interaction coupled to the two different Cooper 
pairs $\phi_1$ and $\phi_2$. In the absence of the SU(2) 
gauge interaction this Lagrangian describes a very 
interesting non-Abelian two-gap superconductor which 
admits two types of non-Abelian Abrikosov vortex \cite{epjb08}.
It has D-type magnetic vortex which has no concentration 
of the condensate at the core and N-type magnetic vortex 
which has a non-trivial profile of the condensate at the core. 
They are described by the non-Abelian topology $\pi_2(S^2)$ 
and  $\pi_1(S^1)$, as well as the Abelian topology 
$\pi_1(S^1)$. And they can carry both integer and fractional magnetic flux \cite{epjb08}. This is because the doublet $\phi$ naturally introduce the non-Abelian structure. This justifies 
the overall U(1) gauge interaction. 

Moreover, in the presence of the doublet $\phi$ it becomes 
natural (and logical) to introduce the interaction between 
$\phi_1$ and $\phi_2$. This justifies the SU(2) gauge 
interaction in (\ref{lag0}). But one might still wonder how 
can one justify the two off-diagonal charged gauge bosons, 
since condensed matters do not appear to have a place 
for them. Actually, in condensed matters the two spin-half 
electrons could form a charged spin-one bound state 
which could be represented by the off-diagonal gauge 
bosons, so that they could play important role in condensed 
matters. In any case, we could always remove the off-diagonal gauge bosons if necessary, as we will see later. With this 
understanding we can say that the above U(2) gauge 
theory could describe a wide class of non-Abelian
condensed matters.

To proceed we express the complex $\phi$ with
the scalar field $\rho$ and the unit doublet $\xi$ by
\begin{gather}
\phi = \frac{1}{\sqrt{2}} \rho~\xi,
~~~(\xi^\dagger \xi = 1),
\end{gather}
and have
\begin{gather}
{\cal L}=-\frac12 (\pd_\mu \rho)^2
- \frac{\rho^2}{2} |{\cal D}_\mu \xi |^2
-\frac{\lambda}{8}\big(\rho^2-\rho_0^2 \big)^2  \nn\\
-\frac14 F_\mn^2 -\frac14 \G_\mn^2,
\label{lag1}
\end{gather}
where $\rho_0=\sqrt{2\mu^2/\lambda}$ is the vacuum
expectation value of the complex doublet field. To simplify
this further we need the Abelian decomposition of
the Lagrangian (\ref{lag0}) \cite{prd80,prl81}.

Consider the SU(2) gauge field $\B_\mu$ first. Let
$(\n_1,\n_2,\n_3=\n)$ be an arbitrary right-handed
orthonormal SU(2) basis, and choose $\n$ to be
the Abelian direction at each space-time point.
Imposing the condition on the gauge potential $\B_\mu$,
\begin{gather}
D_\mu \hn=0,
\label{icon}
\end{gather}
we can project out the restricted potential $\hB_\mu$ which describes the Abelian subdynamics of the non-Abelian 
gauge theory \cite{prd80,prl81}
\begin{gather}
\B_\mu \rightarrow \hB_\mu =\tB_\mu +\tC_\mu, \nn\\
\tB_\mu= B_\mu \n~~(B_\mu=\n \cdot \B_\mu),
~~~\tC_\mu=-\frac{1}{g'} \n\times \pd_\mu \n.
\label{rp}
\end{gather}
Notice that the restricted potential is precisely
the potential which leaves $\n$ invariant under
parallel transport (which makes $\n$ covariantly
constant). Remarkably it has a dual structure,
made of two potentials $\tB_\mu$ and $\tC_\mu$.

With this we obtain the gauge independent Abelian
decomposition of the $SU(2)$ gauge field adding
the valence part $\W_\mu$ which was excluded
by the isometry \cite{prd80,prl81}
\begin{gather}
\B_\mu = \hB_\mu + \W_\mu,
~~~\W_\mu=W_\mu^1 \n_1+W_\mu^2 \n_2.
\label{cdec}
\end{gather}
Under the (infinitesimal) gauge transformation
\begin{gather}
\delta \B_\mu = \frac1{g'}  D_\mu \valpha,
~~~\delta \n = - \valpha \times \n,
\label{gt}
\end{gather}
we have
\begin{gather}
\delta B_\mu = \frac1{g'} \n \cdot \pro_\mu \valpha, \nn\\
\delta \hB_\mu = \frac1{g'} \hD_\mu \valpha,
~~~\delta \W_\mu = -\valpha \times \W_\mu.
\label{cgt}
\end{gather}
This tells that $\hB_\mu$ by itself describes an $SU(2)$
connection which enjoys the full $SU(2)$ gauge degrees
of freedom. Furthermore the valence potential $\W_\mu$
forms a gauge covariant vector field. But what is really
remarkable is that this decomposition is gauge independent.
Once $\n$ is chosen, the decomposition follows
automatically, regardless of the choice of gauge.

The restricted field strength $\hG_\mn$ inherits
the dual structure of $\hB_\mu$, which can also
be described by two Abelian potentials $B_\mu$
and $C_\mu$,
\begin{gather}
\hG_\mn= \pd_\mu \hB_\nu-\pd_\nu \hB_\mu
+ g \hB_\mu \times \hB_\nu =G_\mn' \n, \nn \\
G'_\mn=G_\mn + H_\mn
= \pd_\mu B'_\nu-\pd_\nu B'_\mu,  \nn\\
G_\mn =\pd_\mu B_\nu-\pd_\nu B_\mu, \nn\\
H_\mn = -\frac1{g'} \n \cdot (\pd_\mu \n \times\pd_\nu \n)
=\pd_\mu C_\nu-\pd_\nu C_\mu,  \nn\\
C_\mu = -\frac{2i}{g'} \xi^\dagger \pd_\mu \xi
=-\frac1{g'} \n_1\cdot \pd_\mu \n_2,   \nn\\
B_\mu' = B_\mu+ C_\mu.
\end{gather}
Notice that the potential $C_\mu$ for $H_\mn$ is
determined uniquely up to the $U(1)$ gauge freedom
which leaves $\n$ invariant.

To understand the meaning of $C_\mu$, let
\begin{gather}
\xi =\exp (-i \gamma) \left(\begin{array}{cc}
\sin \frac{\alpha}{2}~\exp (-i \beta) \\
- \cos \frac{\alpha}{2} \end{array} \right), \nn\\
\n=-\xi^\dagger \vsig \xi
=\left(\begin{array}{ccc}
\sin \alpha \cos \beta \\
\sin \alpha \sin \beta \\
\cos \alpha  \end{array} \right).
\label{xi}
\end{gather}
But here we can always put $\gamma=0$ without loss
of generality, because (\ref{lag0}) has the overall U(1)
gauge degrees of freedom. With this we have
\begin{gather}
C_\mu=-\frac1{g'} (1-\cos \alpha) \pd_\mu \beta,  \nn\\
\tC_\mu=-\frac1{g'}~\n\times \pd_\mu \n
= \frac1{g'} \big(\n_1~\sin \alpha~\pd_\mu \beta
-\n_2~\pd_\mu \alpha \big),  \nn\\
\n_1=\left(\begin{array}{ccc}
\cos \alpha \cos \beta \\
\cos \alpha \sin \beta \\
-\sin \alpha  \end{array} \right),
~~~\n_2=\left(\begin{array}{ccc}
- \sin \beta \\ \cos \beta \\
0  \end{array} \right).
\label{monp}
\end{gather}
So when $\n=\hr$, the potential $C_\mu$ describes
the Dirac monopole and $\tC_\mu$ describes the Wu-Yang monopole \cite{wu,prl80}. This justifies us to call $B_\mu$ 
and $C_\mu$ the electric and magnetic potential. Moreover,
this exercise tells that physically the Wu-Yang monopole 
is nothing but the non-Abelian realization of the Dirac monopole. 

Since $\hB_\mu$ has the full SU(2) gauge degrees
of freedom, it inherits all topological characteristics
of the original non-Abelian potential. First, it has
the non-Abelian monopole described by
the $\pi_2(S^2)$ topology of $\n$. Second, it retains
the topologically distinct vacua characterized by
the Hopf invariant $\pi_3(S^2)\simeq\pi_3(S^3)$ of
$\n$ \cite{bpst,plb79}. This means that we can construct
the restricted gauge theory which has the full SU(2)
gauge invariance with the restricted potential $\hB_\mu$.

Moreover, with (\ref{cdec}) we have
\begin{gather}
\G_\mn=\hG_\mn + \hD _\mu \W_\nu - \hD_\nu
\W_\mu + g' \W_\mu \times \W_\nu,   \nn\\
\hD_\mu=\pd_\mu+g' \hB_\mu \times,
\end{gather}
so that the SU(2) Lagrangian is decomposed into
the restricted part and the valence part gauge
independently,
\begin{gather}
{\cal L}_{SU(2)} =-\frac14 \G_\mn^2
=-\frac14 \hG_\mn^2
-\frac14 (\hD_\mu\W_\nu-\hD_\nu\W_\mu)^2 \nn\\
-\frac{g'}{2} \hG_\mn \cdot (\W_\mu \times \W_\nu)
-\frac{g'^2}{4} (\W_\mu \times \W_\nu)^2.
\label{cdec2}
\end{gather}
This is the Abelian decomposition of the SU(2) gauge
theory known as the Cho decomposition, Cho-Duan-Ge
(CDG) decomposition, or  Cho-Faddeev-Niemi (CFN)
decomposition \cite{fadd,shab,zucc,kondo}. In this form,
the theory can be interpreted as the restricted gauge
theory which has the valence part as a gauge covariant
source.

An important advantage of the Abelian decomposition
is that it can actually ``abelianize"  the non-Abelian
gauge theory gauge independently \cite{prd80,prl81}.
Indeed with $W_\mu =(W^1_\mu + i W^2_\mu )/{\sqrt 2}$,
we can put (\ref{cdec2}) in the Abelian form
\begin{gather}
{\cal L}_{SU(2)} = -\frac14 {G'}_\mn^2
-\frac12 |D'_\mu W_\nu-D'_\nu W_\mu|^2  \nn\\
+ ig' G'_\mn W_\mu^*W_\nu
+ \frac{g'^2}{4}(W_\mu^* W_\nu -W_\nu^* W_\mu)^2,  \nn\\
~~~D'_\mu=\pd_\mu+ig' B'_\mu.
\label{adec1}
\end{gather}
One might wonder how the non-Abelian structure
disappears in this Abelianization. Actually the non-Abelian
structure has not disappeared but hidden. To see this
notice that the potential $B'_\mu$ in the Abelian formalism
is dual, given by the sum of the electric and magnetic
potentials $B_\mu$ and $C_\mu$. Clearly $C_\mu$
represents the topological degrees of the non-Abelian
symmetry which does not exist in the naive Abelianization
that one obtains by fixing the gauge, choosing 
$\n=(0,0,1)$ \cite{prd80,prl81}.

The Abelian decomposition has played a crucial role in QCD
for us to prove the monopole condensation create the mass 
gap to generate the color confinement \cite{prd13,epjc19}.
But now it must be clear that it could also play an important 
role in condensed matter physics, because it teaches us 
how the Abelian gauge theory can be embedded in non-Abelian gauge theory and how the electromagnetic interaction can 
arise from the non-Abelian gauge interaction.

With the Abelian decomposition, the Lagrangian (1)
has two Abelian potentials, $A_\mu$ and $B'_\mu$,
and we need to clarify the physical meaning of them.
Clearly we can identify $A_\mu$ as the electromagnetic 
potential if the two condensates $\phi_1$ and $\phi_2$ 
have the same charge, but $B'_\mu$ as the electromagnetic potential if they have opposite charge. However, since this 
depends on what kind of materials we have at hand, it is 
better to identify the electromagnetic potential to be a linear
combination of two potentials.

For this reason we define the electromagnetic potential
$A_\mu^{\rm (em)}$ and another potential $X_\mu$ with
the mixing angle $\theta$ by
\begin{gather}
\left( \begin{array}{cc} A_\mu^{\rm (em)} \\
X_\mu  \end{array} \right)
=\frac{1}{\sqrt{g^2 + g'^2}} \left(\begin{array}{cc} g' & g \\
-g & g' \end{array} \right)
\left( \begin{array}{cc} A_\mu \\ B'_\mu
\end{array} \right)  \nn\\
= \left(\begin{array}{cc}
\cos\theta & \sin\theta  \\
-\sin\theta & \cos\theta  \end{array} \right)
\left( \begin{array}{cc} A_\mu \\ B_\mu'
\end{array} \right).
\label{mix}
\end{gather}
Notice that when $\theta$ is zero (i.e., with g=0)
$A_\mu$ becomes $A_\mu^{\rm (em)}$, but when
$\theta$ becomes $\pi/2$ (i.e., with $g'=0$), $B'_\mu$
describes the electromagnetic potential. But in general
when $gg' \ne 0$, $A_\mu^{\rm (em)}$ is given by
a linear combination of $A_\mu$ and $B'_\mu$.

From this we have the identity
\begin{gather}
D_\mu \xi= -i\frac{g}{2} (B_\mu' \n +\W_\mu) \cdot \vsig~\xi,  \nn\\
|D_\mu \xi|^2 =\frac{g'^2}{4} ({B'}_\mu^2
+ \W_\mu^2),  \nn\\
|{\cal D}_\mu \xi|^2 =\frac{g^2+g'^2}{8} X_\mu^2 
+\frac{g'^2}{4} \W_\mu^2.
\end{gather}
With this we can remove the doublet $\phi$ completely
from the Lagrangian (\ref{lag0}) and abelianize it
gauge independently in the following form
\begin{gather}
{\cal L} = -\frac12 (\pd_\mu \rho)^2
-\frac{\lam}{8}\big(\rho^2-\rho_0^2 \big)^2 \nn\\
-\frac14 {F_\mn^{\rm (em)}}^2
-\frac14 X_\mn^2
-\frac{\rho^2}{4} \big(g'^2 W_\mu^*W_\mu
+\frac{g^2+g'^2}{2} X_\mu^2 \big)   \nn\\
-\frac12 \big|(D_\mu^{\rm (em)}
+ie\frac{g'}{g} X_\mu)W_\nu -(D_\nu^{\rm (em)}
+ie\frac{g'}{g} X_\nu)W_\mu)\big|^2  \nn\\
+ie (F_\mn^{\rm (em)}
+ \frac{g'}{g}  X_\mn) W_\mu^* W_\nu \nn\\
+ \frac{g'^2}{4}(W_\mu^* W_\nu - W_\nu^* W_\mu)^2,
\label{adec2}
\end{gather}
where $D_\mu^{\rm (em)}=\pd_\mu+ieA_\mu^{\rm (em)}$
and $e$ is the electric charge
\begin{gather}
e=\frac{gg'}{\sqrt{g^2+g'^2}}=g' \sin\theta
=g \cos\theta.
\end{gather}
In this form the Lagrangian describes QED coupled to two 
gauge bosons, the charged $W_\mu$ boson and the neutral 
$X_\mu$ boson, which acquire the mass
\begin{gather}
M_W=\frac{g'}{2}~\rho_0,
~~~~M_X=\frac{\sqrt{g^2+g'^2}}{2}~\rho_0,
\label{mass}
\end{gather}
through the Higgs mechanism. The popular interpretation 
of this is that the gauge bosons acquire the mass by 
``the  spontaneous symmetry breaking" of U(2) down to
the electromagnetic U(1) through the Higgs vacuum.  
But we emphasize that here the gauge bosons acquire 
the mass without any symmetry breaking, spontaneous 
or not. All we have to do is the reparametrize the fields
which does not involve any symmetry breaking.

Notice that we can easily switch off the X boson if necessary. 
In this case the Lagrangian reduces to
\begin{gather}
{\cal L}_1= -\frac{1}{2}(\pd_\mu \rho)^2
-\frac{\lambda}{8}\big(\rho^2-\rho_0^2 \big)^2   \nn\\
-\frac14 {F_\mn^{\rm (em)}}^2
-\frac12 |D_\mu^{\rm (em)} W_\nu
-D_\nu^{\rm (em)} W_\mu|^2   \nn\\
-\frac{g'^2}{4}\rho^2 W_\mu^*W_\mu
+ie F_\mn^{\rm (em)} W_\mu^* W_\nu  \nn\\
+ \frac{g'^2}{4}(W_\mu^* W_\nu - W_\nu^* W_\mu)^2.
\label{wlag}
\end{gather}
This describes QED coupled to a charged vector field
$W_\mu$ as the source whose mass comes from
the Higgs mechanism. Moreover, when we swtch off 
$F_\mn^{\rm (em)}$ and $W_\mu$, the Lagrangian 
reduces to
\begin{gather}
{\cal L}_2 = -\frac{1}{2}(\pd_\mu \rho)^2
-\frac{\lambda}{8}\big(\rho^2-\rho_0^2 \big)^2   \nn\\
-\frac14 X_\mn^2 -\frac{g^2+g'^2}{8} \rho^2 X_\mu^2.
\end{gather}
This shows that the Lagrangian (\ref{lag0}) can describe 
a variety of non-Abelian condensed matters.

\section{Monopole Configuration in Two-Gap
Condensed Matters}

It is well known that QED has the Dirac monopole,
and SU(2) gauge theory has the 'tHooft-Polyakov
monopole \cite{dirac,thooft}. If so, it is quite likely that
the above Lagrangian has a monopole-like topological
object. In this section we show that indeed (\ref{lag0})
has a very interesting monopole which can be generalized
to a dyon. To show this we choose the following ansatz in
the spherical coordinates $(t,r,\theta,\varphi)$,
\begin{gather}
\rho =\rho(r),~~~~\xi
=i\left(\begin{array}{cc} \sin (\theta/2)~e^{-i\varphi} \\
- \cos(\theta/2) \end{array} \right),   \nn\\
A_\mu =\frac{1}{g} A(r) \pd_\mu t
-\frac{1}{g}(1-\cos\theta) \pd_\mu \varphi,  \nn\\
\B_\mu= \frac{1}{g'} B(r)\pd_\mu t~\hr
+\frac{1}{g'}(f(r)-1)~\hr \times \pd_\mu \hr,  \nn\\
\n=-\xi^\dagger \vsig~\xi =\hr.
\label{ans1}
\end{gather}
It has the following features. First, when $A(r)=B(r)=0$,
$A_\mu$ describes the Dirac-type Abelian monopole and
$\B_\mu$ describes the 'tHooft-Polyakov monopole.
So the ansatz is a hybrid   between Dirac and
'tHooft-Polyakov. Second, it has the Coulomic part $A(r)$
and $B(r)$ which could add the electric charge to the monopole
to make it a dyon. To see this, notice that in terms of
the physical fields the ansatz (\ref{ans1}) can be expressed
by
\begin{gather}
A_\mu^{\rm (em)} = e\Big( \frac{A(r)}{g^2}
+ \frac{B(r)}{g'^2}  \Big) \pd_\mu t
-\frac{1}{e}(1-\cos\theta) \pd_\mu \varphi,  \nn \\
X_\mu = \frac{e}{gg'}\big(B(r)-A(r) \big) \pd_\mu t,  \nn\\
W_\mu=\dfrac{i}{g'}\frac{f(r)}{\sqrt2} e^{i\varphi}
(\pd_\mu \theta +i \sin\theta \pd_\mu \varphi).
\label{ans2}
\end{gather}
This clearly shows that the ansatz is for a real
electromagnetic dyon.

\begin{figure}
\includegraphics[height=4.5cm, width=7.5cm]{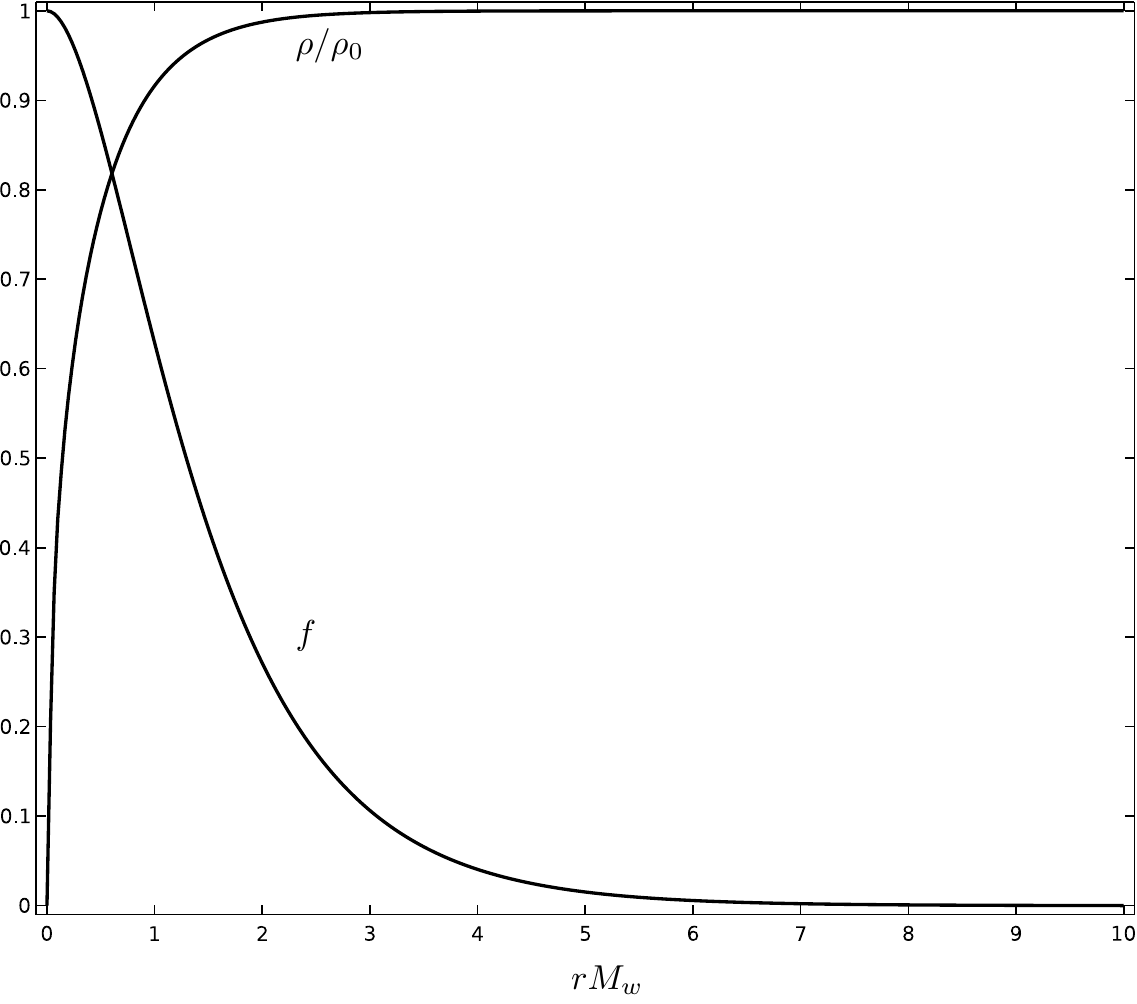}
\caption{\label{cmm} The singular monopole solution
with W-boson and Higgs scalar dressing in two-gap
superconductor.}
\end{figure}

The ansatz reduces the equations of motion to
\begin{gather}
\ddot{\rho}+\frac{2}{r} \dot{\rho}-\frac{f^2}{2r^2}\rho
+\frac14 (B-A)^2\rho
=\frac {\lambda}{2}\big(\rho^2-\rho_0^2 \big)\rho , \nn\\
\ddot{f}-\frac{f^2-1}{r^2}f=\big(\frac{g'^2}{4}\rho^2
- B^2\big)f, \nn\\
\ddot{A} +\frac{2}{r} \dot{A}
=-\frac{g^2}{4} \rho^2 (B-A),    \nn\\
\ddot{B}+\frac{2}{r}\dot{B}-\frac{2f^2}{r^2}B
=\frac{g'^2}{4}\rho^2(B-A).
\label{deq1}
\end{gather}
Obviously this has a trivial solution
\begin{gather}
\rho=\rho_0=\sqrt{2\mu^2/\lambda},~~~f=0,
~~~A=B=0,
\end{gather}
which describes the Dirac type point monopole
\begin{gather}
A_\mu^{\rm (em)}=-\frac{1}{e}(1
-\cos \theta) \pd_\mu \varphi.
\label{smon}
\end{gather}
This monopole has two remarkable features. First, this
is not the Dirac's monopole. It has the electric charge
$4\pi/e$, not $2\pi/e$ \cite{plb97}. Second, this monopole
naturally admits a non-trivial dressing of W boson.
Indeed we can integrate (\ref{deq1}) with $A=B=0$ and
the boundary conditions
\begin{gather}
\rho(0)=0,~~~\rho(\infty)=\rho_0,   \nn\\
f(0)=1,~~~f(\infty)=0.
\label{mbcon}
\end{gather}
This gives the dressed monopole solution shown in 
Fig. \ref{cmm}. Clearly this can be viewed as a hybrid 
between the Dirac monopole and the 'tHooft-Polyakov 
monopole.

Moreover, we can extend it to a dyon solution with the following boundary conditions
\begin{gather}
\rho(0)=0,~~f(0)=1,~~A(0)=a_0,~~B(0)=0, \nn\\
\rho(\infty)=\rho_0,~f(\infty)=0,~A(\infty)=B(\infty)=A_0,
\label{dbcon}
\end{gather}
and can show that the equation (\ref{deq1}) admits
a family of solutions labeled by the real parameter $A_0$
lying in the range
\begin{gather}
0 \leq A_0 < {\rm min} ~\Big(e\rho_0,
~\frac{g'}{2}\rho_0\Big).
\label{boundA}
\end{gather}
Near the origin it has the following behavior,
\begin{gather}
\rho \simeq \alpha_1 r^{\delta_-},
~~~~~f \simeq 1+ \beta_1  r^2,  \nn \\
A \simeq a_0 + a_1 r^{2\delta_+},
~~~~~B \simeq b_1 r,
\label{origin}
\end{gather}
where $\delta_{\pm} =(\sqrt{3} \pm 1)/2$.
Asymptotically we have
\begin{gather}
\rho \simeq \rho_0 +\rho_1\frac{\exp(-\sqrt{2}\mu r)}{r},  
~~~f \simeq  f_1 \exp(-\omega  r),  \nn\\
A \simeq B +A_1 \frac{\exp(-\nu r)}{r},
~~~B \simeq A_0 +\frac{B_1}{r},
\label{infty}
\end{gather}
where $\omega=\sqrt{(g\rho_0)^2/4 -A_0^2}$,
and $\nu=\sqrt{(g^2 +g'^2)}\rho_0/2$.

\begin{figure}
\includegraphics[height=4.5cm, width=7.5cm]{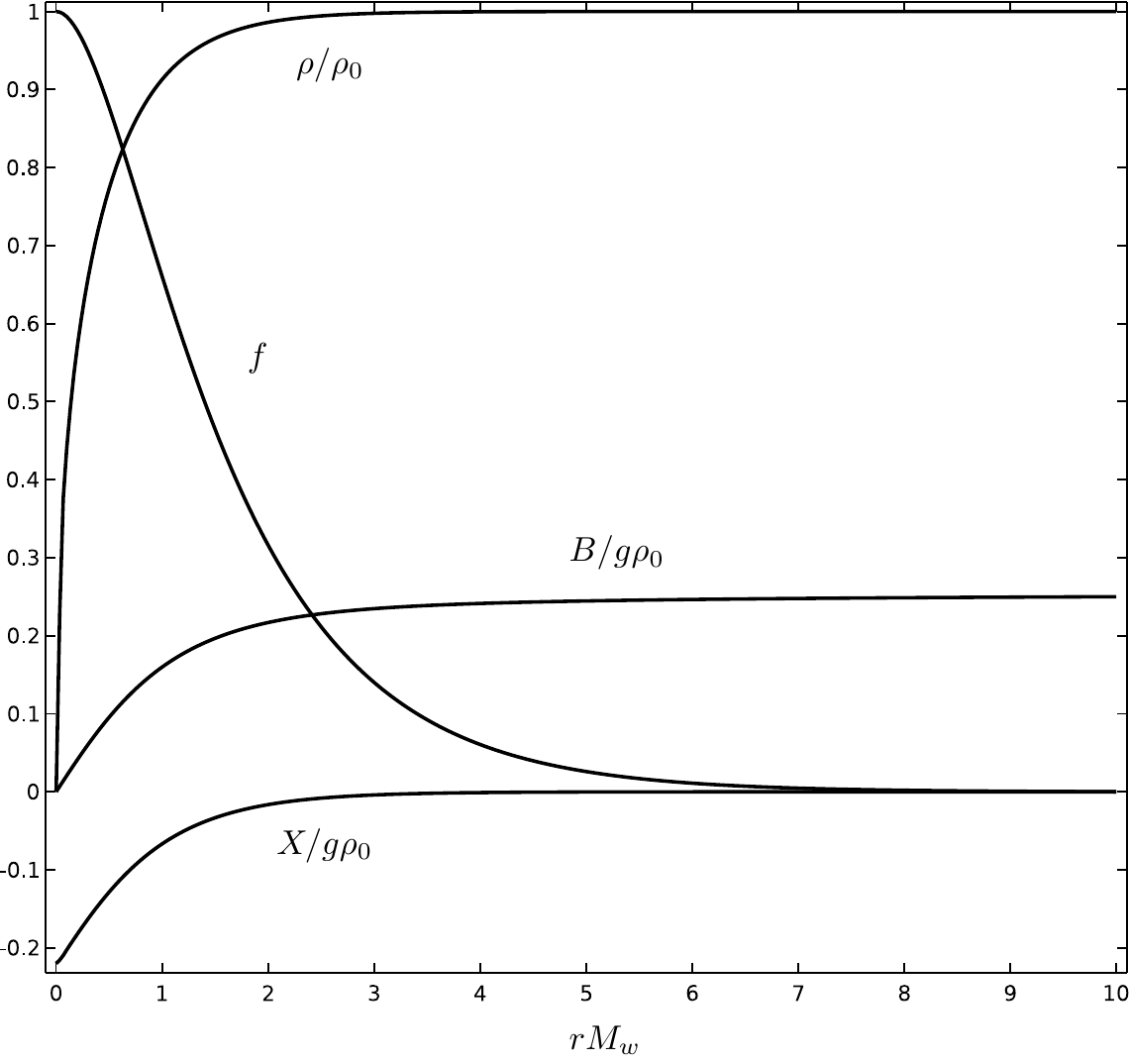}
\caption{\label{cmd} The singular dyon solution
with W boson, X boson, and Higgs scaler dressing in
two-gap superconductor. Here $X=B-A$ and we have
chosen $A(\infty)=g\rho_0/4$.}
\end{figure}

This tells that $M_H$, $\sqrt{1-(A_0/M_W)^2}~M_W$,
and $M_X$ determine the exponential damping of 
the Higgs boson, $W$ boson, and $X$ boson to their 
vacuum expectation values asymptotically. The dyon 
solution is shown in Fig. \ref{cmd}.

The dyon has the following electromagnetic charges
\begin{gather}
q_e=-\frac{8\pi}{e}\sin^2\theta \Int_0^\infty f^2 B dr
=\frac{4\pi}{e} A_1, \nn\\
q_m = \dfrac{4\pi}{e}.
\label{mcharge}
\end{gather}
Moreover, the dyon equation (\ref{deq1}) is invariant under 
the reflection 
\begin{gather}
A \rightarrow -A,~~~~~ B\rightarrow -B. 
\end{gather}
This means that, for a given magnetic charge, there are 
two dyon solutions which carry opposite electric charges 
$\pm q_e$. We can also have the anti-monopole or 
in general anti-dyon solution, the charge conjugate state 
of the dyon which has the magnetic charge $q_m=-4\pi/e$.

\section{Regularization of Monopole in Two-Gap Condensed Matters}

As we have pointed out the above monopole is a hybrid
between the Dirac and 'tHooft-Polyakov monopoles,
so that it has an infinite energy. Of course classically there
is nothing wrong with this. Nevertheless one might wonder
if we can regularize the monopole to a finite energy.
One might think there is no reason to regularize the monopole,
because in real condensed matters we have a natural cut-off 
given by the atomic size which can make the monopole energy 
finite. This might be so, but from the theoretical point of view, 
it would be nice to make the monopole energy finite without 
any cut-off. Now, we show that the quantum correction at short distance could make the monopole energy finite.

\begin{figure}
\includegraphics[height=4.5cm, width=8cm]{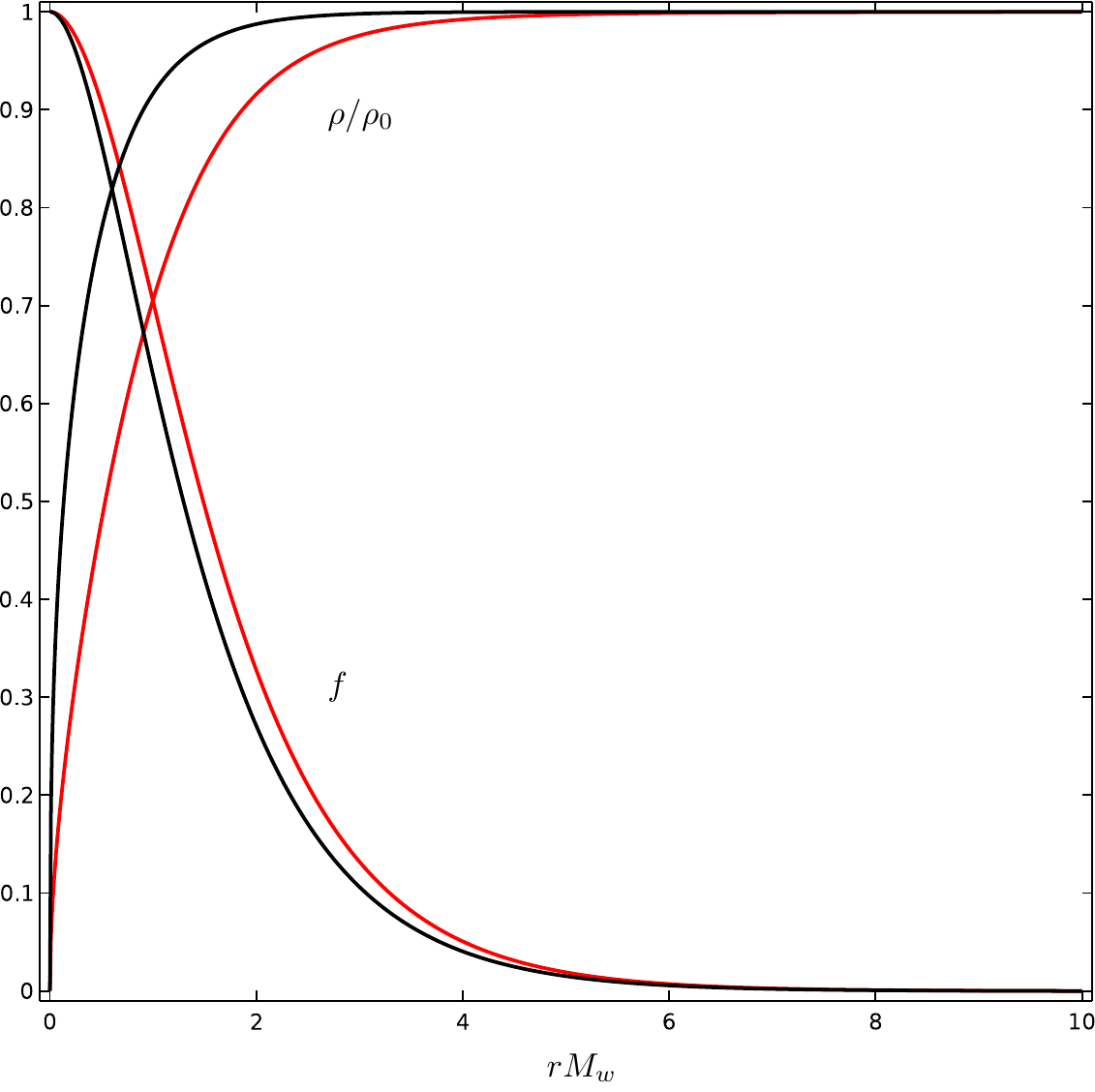}
\caption{\label{fem} The finite energy monopole solution
(the red curves) regularized by the real electromagnetic
permittivity $\eps_1=(\rho/\rho_0)^6$ with the W boson
and Higgs field dressing. For comparison we plot
the singular monopole solution shown in Fig. \ref{cmm}
in black curves.}
\end{figure}

Consider the following effective Lagrangian
\begin{gather}
{\cal \bar L}= -\frac12 (\pd_\mu \rho)^2
-\frac{\lambda}{8}\big(\rho^2-\rho_0^2 \big)^2
-\frac14 \eps(\rho) {F_\mn^{\rm (em)}}^2   \nn\\
-\frac14 X_\mn^2
-\frac{\rho^2}{4} \big(g'^2 W_\mu^*W_\mu
+\frac{g^2+g'^2}{2} X_\mu^2 \big)   \nn\\
-\frac12 \big|(D_\mu^{\rm (em)}
+ie\frac{g'}{g} X_\mu)W_\nu -(D_\nu^{\rm (em)}
+ie\frac{g'}{g} X_\nu)W_\mu)\big|^2  \nn\\
+ie (F_\mn^{\rm (em)}
+ \frac{g'}{g}  X_\mn) W_\mu^* W_\nu \nn\\
+ \frac{g'^2}{4}(W_\mu^* W_\nu - W_\nu^* W_\mu)^2,
\label{lag2}
\end{gather}
where $\eps$ is the real non-vacuum electromagnetic
permittivity of the condensed matter. It retains all symmetries 
of the Lagrangian  (\ref{adec2}). The only difference is that 
here we have $\eps$ in front of $F_\mn^{\rm (em)}$ which 
effectively changes the electromagnetic coupling constant 
$e$ to a running coupling $e/{\sqrt \epsilon}$. 

\begin{figure}
\includegraphics[height=4.5cm, width=8cm]{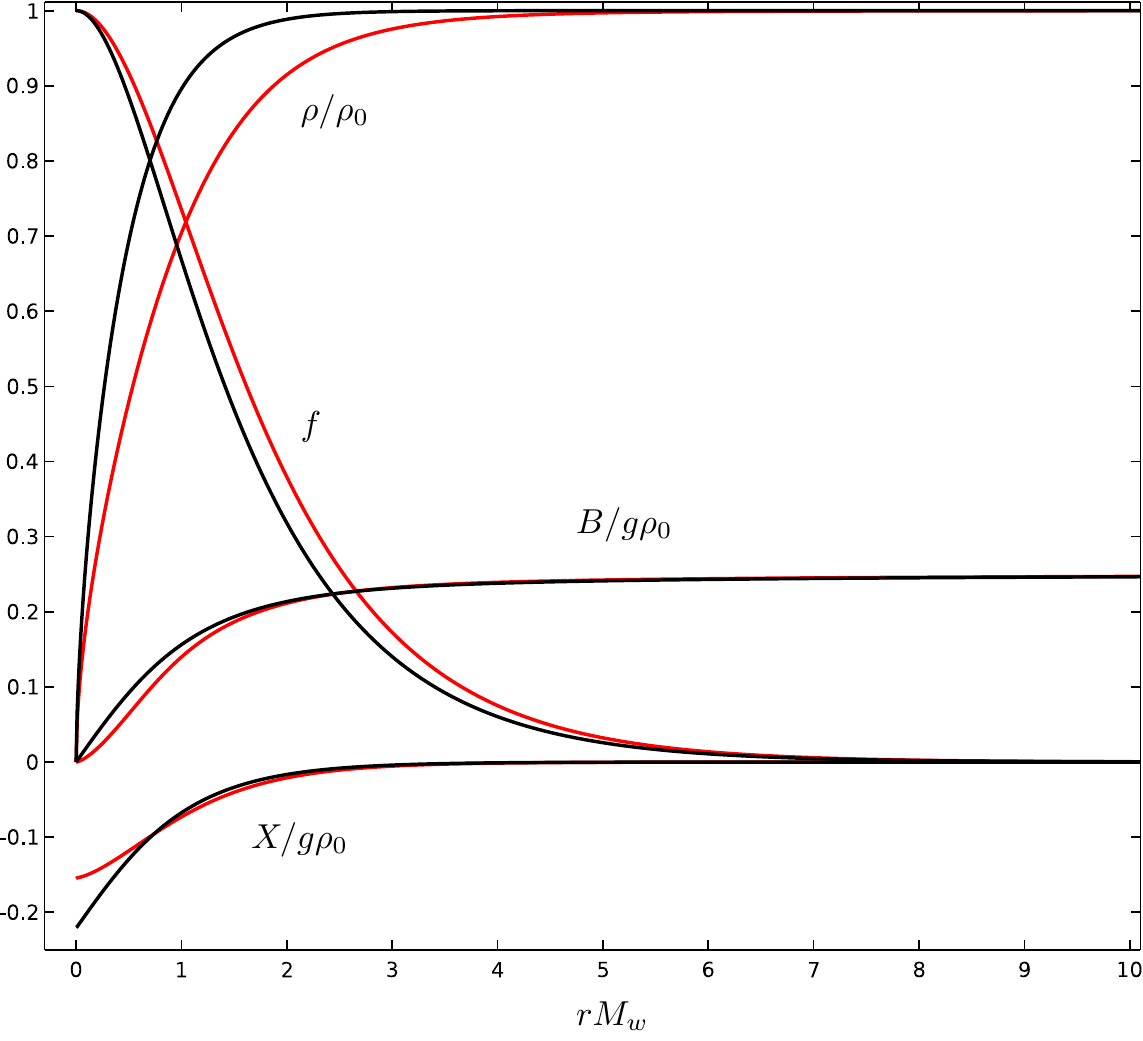}
\caption{\label{fed} The finite energy dyon solution
(the red curves) regularized by the real electromagnetic
permittivity $\eps=(\rho/\rho_0)^6$ with the W boson,
X boson, and Higgs field dressing. For comparison we plot
the singular dyon solution shown in Fig. \ref{cmd}
in black curves.}
\end{figure}

This, with the ansatz (\ref{ans2}), gives the following 
equation of motion
\begin{gather}
\ddot \rho + \frac{2}{r}\dot \rho-\frac{f^2}{2r^2} \rho
=\frac{\lambda}{2} (\rho^2- \rho_0^2) \rho
-\frac{1}{4} (B-A)^2 \rho    \nn\\
+\frac{\eps'}{2}\Big(\frac{1}{e^2 r^4}
-e^2(\frac{\dot A}{g^2}+\frac{\dot B}{g'^2})^2 \Big),  \nn\\
\ddot{f}-\frac{f^2-1}{r^2}f
=\big(\frac{g'^2}{4} \rho^2 - B^2\big) f, \nn\\
\ddot A + \frac{2}{r} \dot A
+e^2 \frac{\eps'}{\eps}\dot \rho \big(\frac{\dot A}{g^2}
+\frac{\dot B}{g'^2}\big)
+\frac{2 e^2}{g'^2} \big(1-\frac{1}{\eps}\big)
\frac{f^2}{r^2} B  \nn\\
=-\frac{g^2}{4} \rho^2 (B-A),  \nn\\
\ddot B +\frac{2}{r}\dot B
+e^2 \frac{\eps'}{\eps} \dot \rho \big(\frac{\dot A}{g^2}
+\frac{\dot B}{g'^2}\big)
-\frac{2e^2}{g'^2} \big(\frac{g'^2}{g^2}
+\frac{1}{\eps} \big) \frac{f^2}{r^2} B \nn\\
=\frac{g'^2}{4}(B-A) \rho^2.
\label{fedeq2}
\end{gather}
Notice that if we switch off the W and X bosons, the above Lagrangian reduces to the Lagrangian (\ref{dlag}), and 
the equations of motion (with $f=A=B=0$) reduces to 
(\ref{mdeq}). This confirms that the above effective 
Lagrangian could be interpreted as a non-Abelian 
generalization of (\ref{dlag}). 

To integrate it out and find a finite energy solution,
we have to choose a proper boundary condition.
To do this notice that with
\begin{gather}
\eps =\eps_0 +\eps_1,
~~~\eps_0= \frac{g'^2}{g^2+g'^2} ,  
~~~\eps_1 \simeq \Big(\frac{\rho}{\rho_0}\Big)^n,  
\label{epbc}
\end{gather}  
the energy is expressed by
\begin{gather}
E=4\pi \int_0^\infty dr \bigg\{\frac{1}{2e^2 r^2}
\Big(\eps_0 (f^2-1)^2+ \eps_1 \Big)
+\frac12 (r\dot \rho)^2   \nn\\
+\frac{\lambda r^2}{8}\big(\rho^2-\rho_0^2 \big)^2
+\frac1{g^2} \dot f^2 +\frac14 f^2 \rho^2 
+ \frac{f^2 A^2}{g^2}  \nn\\
+\frac{r^2}{8}(A-B)^2\rho^2
+\frac{r^2(\dot A-\dot B)^2}{2(g^2+g'^2)}  \nn\\
+\frac{\eps~e^2 r^2}{2} \big(\frac{\dot A}{g^2}
+\frac{\dot B}{g'^2} \big)^2  \bigg\}.
\label{fede2}
\end{gather}
This tells that we can make the energy finite imposing 
the boundary condition $f(0)=1$, when $n >2$. In other 
words we can regularize the monopole with the real 
electromagnetic permittivity $\eps$ making $\eps(0)$ 
finite.

To see that this regularization works, consider the monopole
solution first. With $A=B=0$ we can integrate (\ref{fedeq2})
with the boundary condition $f(0)=1$, and obtain the monopole
solution shown in Fig. \ref{fem}. We can generalize this
to the dyon solution solving (\ref{fedeq2}), and obtain
the finite energy dyon solution shown in Fig. \ref{fed}.
This confirms that we can indeed regularize monopole and
dyon solutions with the real electromagnetic permittivity.
Notice here that a non-vanishing value of $\eps$ at the origin
plays an essential role to yield the finite energy solution.

One might ask if we can have a regularized monopole 
solution without $f$ (without the W boson). This is a relevant question because many of real condensed matters may 
not have the W boson (the charged spin-one two-electron 
bound state). Clearly this is possible, because with 
$f=A=B=0$ the equation (\ref{fedeq2}) reduces to 
(\ref{mdeq}) which has the regularized Dirac monopole 
solution shown in Fig. \ref{femh}. The only thing is that 
here we must choose $\eps=(\rho/\rho_0)^n$ to make 
the energy (\ref{fede2}) finite. This tells that we can actually regularize the monopole without the W boson with the real electromagnetic permittivity. Moreover, this reconfirms that 
the above monopole is the non-Abelian generalization 
of the Dirac monopole. 

One might try to generalize this to a dyon solution. But
with $f=0$, we find that (\ref{fedeq2}) has no solution
with non-trivial $A$ and $B$. The results in this section 
confirm that we can regularize the monopole in condensed 
matters replacing the vacuum electric permittivity with 
a real electric permittivity, even in the absence of 
the W  boson. Clearly the results enhance the possibility 
for a monopole to exist in real superconductors.

\section{Comparison with Electroweak Monopole}

One might have noticed that the monopole and dyon
solutions discussed in Sections III and IV look very similar
to the electroweak monopole and dyon solutions in
the standard model known as the Cho-Maison monopole
and dyon \cite{plb97,yang,epjc15,mpla16,ellis,bpscm}.
In fact we can easily confirm that mathematically they are 
identical, so that formally there is exactly one to one 
correspondence between the Lagrangian (\ref{lag0}) and 
the Weinberg-Salam Lagrangian. The mixing angle
$\theta$ in (\ref{mix}) corresponds to the Weinberg
angle, the W boson and X boson correspond to
the W boson and Z boson in the standard model.
So, the above monopole and dyon correspond exactly
to the electroweak monopole and dyon in the standard
model \cite{plb97,yang,epjc15,mpla16,ellis,bpscm}.

From the physical point of view, however, we emphasize
that the two Lagrangians describe completely different
physics. The standard model which unifies
the electromagnetic interaction with the weak interaction
is a fundamental theory of nature. The coupling
constants $g$ and $g'$ in the standard model represent
the fundamental constants which determine the electric 
charge and the Weinberg angle. Moreover, the Higgs particle,
W boson, and Z bosons are the elementary particles
of nature. And the Higgs vacuum $\rho_0$ sets
the electroweak scale of the order of 100 GeV.

On the other hand, our Lagrangian (\ref{lag0}) is an effective Lagrangian describing a low energy physics of condensed 
matters, not a fundamental interaction of nature. So here 
the two coupling constants and the mixing angle in general
depend on the type of the condensed matter, so that
they should not be viewed as fundamental constants
of physics. Moreover, all fields here (except the photon)
should be regarded as not elementary but emergent fields,
so that the W and X bosons should be interpreted as
composite (pseudo) particles. In particular, the Higgs
doublet here describes the two condensates of the non-Abelian condensed matters, which has a typical energy scale of 
several meV.

Consequently, the two monopoles have totally different
meanings. The electroweak monopole is a fundamental
particle which exists in the standard model. So, when
discovered, the monopole will be viewed as the first
absolutely stable topological elementary particle in the history
of physics. On the other hand, the above monopole in
condensed matters is not an elementary particle, but 
an emergent particle existing in condensed matters.
Hence, it may not be stable, even though it is topological.

This can be understood as follows. Consider the Abrikosov 
vortex. We can create it applying magnetic field on 
the superconductor. But it is not fundamental nor stable, 
although it is topological. When we switch off the magnetic 
field, it disappears. The monopole in condensed matters 
here should be similar. We could possibly create it imposing 
the monopole topology by brute force with an external 
magnetic field, but when we switch off the magnetic field, 
it probably will disappear.

Another difference is the mass and size of the monopole.
The mass of the electroweak monopole is expected to be
about 4 to 10 TeV, although the mass of the other electroweak
particles (W boson, Z boson, and Higgs particle) is of
the order of 100 GeV. This is because basically the mass
of the monopole comes from the same Higgs mechanism
which makes the W boson massive, except that the coupling
is magnetic (not $e$ but $4\pi/e$). This means that the mass
of the monopole should be heavier than that of the W boson
by the factor $1/\alpha$. This makes the monopole mass
hundred times heavier than the W boson mass, around
10 TeV \cite{epjc15,mpla16,ellis,bpscm}.

The same logic applies to the monopole in condensed matter.
It is known that the scale of coherence length of ordinary superconductors is roughly of the order of (the inverse of) 
meV, meaning that the mass of the W boson, X boson, 
and Higgs scalar in condensed matters are of the order of 
meV. This implies that the mass of the monopole in condensed
matters should be roughly 100 meV, namely, $1/\alpha$ times heavier than the mass of the Higgs scalar in superconductors.
As for the size, the size of the electroweak monopole
is set by the electroweak scale, the inverse of
100 GeV \cite{plb97,epjc15}, while the size of the monopole 
in condensed matters should be of the order of inverse of 
$O(1)$ meV. This tells that the two monopoles are very 
different, in energy scale by the factor of $10^{13}$, and in 
volume scale by the factor of $10^{39}$. However, it is 
a remarkable beauty that the mathematically same theory 
can describe totally different physics.

\section{Discussions}

In this paper we have discussed the Abelian and non-Abelian monopoles which could exist in condensed matters. In 
particular, we have shown how the non-vacuum electromagnetic permittivity could regularize the Dirac monopole in dielectric condensed matters, and how this regularization could allow 
the monopole with mass around hundreds meV in ordinary condensed matters. Moreover, we have shown how we can generalize this Abelian monopole to a non-Abelian monopole 
in two-gap condensed matters. 

In doing this we have shown how a non-Abelian structure 
can emerge in multi-component condensed matters, 
and how it can be reduced to an Abelian structure with 
the Abelian decomposition. The necessary condition 
for the non-Abelian structure in condensed matter is 
the existence of two or more independent bases, such as 
multi-gap or multi-component states, which can be treated 
as a non-Abelian multiplet. Once this requirement is fulfilled, 
we can naturally introduce the non-Abelian gauge interaction 
to the condensed matters. This condition is naturally satisfied 
in a wide class of condensed matters, e.g., two-gap superconductors, two-component Bose-Einstein condensates, 
and cold atoms with two dark states, etc.

Does this mean that we can have the monopole in condensed matters? Definitely yes. But as we have emphasized in 
Section VI, these monopoles should be regarded as 
emergent objects, not elementary objects. So it is unlikely 
the monopole exist in condensed matters naturally. Just as 
we can create the Abrikosov vortex in type II superconductor, 
we must create them by brute force. Experimentally, this 
may not be easy, but our analysis tells that it is not impossible.   
We hope that our work could stimulate the search for 
the monopole in condensed matters.

At this point one might wonder if the Lagrangian (\ref{lag0})
could also describe a two-gap superconductor. This is 
a very interesting question. As it is, the Lagrangian is not
likely to describe the superconductor, because the photon 
remains massless in (\ref{lag0}). On the other hand we 
have already noticed that, when we remove the SU(2) 
gauge interaction in the Lagrangian (\ref{lag0}), it describes 
a generalized Landau-Ginzburg theory of two-gap 
superconductor which has many interesting features \cite{epjb08}. This strongly implies that the Lagrangian, with some 
modification, has a potential to describe a two-gap 
superconductor. But we need more discussion to show 
exactly under what condition (\ref{lag0}) could describe 
a realistic two-gap superconductor. This matter will be 
discussed in a separate publication.   

{\bf ACKNOWLEDGEMENT}

~~~YKB is supported by the NRF Grant No. 2019-R1H1A2-079920 funded by the Korean government. YMC is supported in part 
by the NRF Grant No. 2018-R1D1A1B0-7045163 funded by 
the Korean government and by Center for Quantum Spacetime, 
Sogang University, Korea. TYS and LPZ are supported by 
the National Natural Science Foundation of China (Grant 
11805242 and Grant 11304062).


\begin{references}
\bibitem{kelvin} L. Kelvin, Trans. Roy. Soc. (Edington) {\bf 25},
217 (1868); P.G. Tait, Scientific Papers, {\bf 1}, 136 (1911).
\bibitem{dirac} P.A.M. Dirac, Phys. Rev. {\bf 74},
817 (1948).
\bibitem{wu} T.T. Wu and C.N. Yang, in {\it Properties
of Matter under Unusual Conditions}, edited by H. Mark
and S. Fernbach (Interscience, New York) 1969;
Phys. Rev. {\bf D12}, 3845 (1975).
\bibitem{thooft} G. 'tHooft, Nucl. Phys. {\bf B79}, 276 (1974);
A. Polyakov, JETP Lett. {\bf 20}, 194 (1974).
\bibitem{julia} B. Julia and A. Zee, Phys. Rev. {\bf D11}, 
2227 (1975); M.K. Prasad and C.M. Sommerfield,
Phys. Rev. Lett. {\bf 35}, 760 (1975).

\bibitem{prl80} Y.M. Cho, Phys. Rev. Lett. {\bf 44},
1115 (1980).
\bibitem{dokos} C.P. Dokos and T.N. Tomaras,
Phys. Rev. {\bf D21}, 2940 (1980).
\bibitem{cab} B. Cabrera, Phys.  Rev. Lett.  {\bf 48}, 1378
(1982).
\bibitem{medal} B. Acharya et al. (MoEDAL Collaboration),
Phys. Rev. Lett. {\bf118}, 061801 (2017); Phys. Rev. Lett.
{\bf 123}, 021802 (2019). 
\bibitem{atlas} ATLAS Collaboration, Euro. Phys. J. {\bf C75}, 
362 (2015); Phys. Rev. Lett. {\bf 124}, 031802 (2020). 

\bibitem{icecube} R. Abbasi et al. (IceCube Collaboration)
Phys. Rev. {\bf D87}, 022001 (2013); M. Aartsen et al.
(IceCube Collaboration), Eur. Phys. J. {\bf C74}, 2938
(2014); S. Adri{\'a}n-Martinez et al. (ANTARES Collaboration),
Astropart. Phys. {\bf 35}, 634 (2012).

\bibitem{plb97} Y.M. Cho and D. Maison, Phys. Lett.
{\bf B391}, 360 (1997).
\bibitem{yang} Yisong Yang, Proc. Roy. Soc. {\bf A454},
155 (1998). See also Yisong Yang, {\it Solitons in Field
Theory and Nonlinear Analysis} (Springer Monographs
in Mathematics), (Springer-Verlag) 2001.
\bibitem{epjc15} Kyoungtae Kimm, J.H. Yoon, and
Y.M. Cho, Eur. Phys. J. {\bf C75}, 67 (2015).

\bibitem{abri} A. Abrokosov, JETP (sov. Phys.) {\bf 5},
1173 (1957); H. Nielsen and P. Olesen, Nucl. Phys.
{\bf B61}, 45 (1973).
\bibitem{prb05} Y.M. Cho, Phys. Rev. {\bf B72}, 212516
(2005).
\bibitem{pra05} Y.M. Cho, Hyojoong Khim, and
Pengming Zhang, Phys. Rev. {\bf A72}, 063603 (2005).
\bibitem{epjb08} Y.M. Cho and P.M. Zhang, Euro Phys. J.
{\bf B65}, 155 (2008); P. Forgacs, S. Reullion, and M. Volkov,
Phys. Rev. Lett. {\bf 96}, 041601 (2006).

\bibitem{cmm1} C. Castelnovo, R. Moessner, and
S.L. Sondhi, Nature {\bf 451}, 42 (2008); S.T. Bramwell,
S.R. Giblin, S. Calder, R. Aldus, D. Prabhakaran, T. Fennel,
Nature {\bf 461}, 956 (2009).
\bibitem{cmm2} V. Pietila and M. Mottonen, Phys. Rev. Lett.
{\bf 103}, 030401 (2009); M. Ray, E. Ruokokoski, S, Kandal,
M. Mollonen, D. Hall, Nature {\bf 505}, 657 (2014);
\bibitem{cmm3} E. Yakaboylu, A. Deuchert, and
M. Lemeshko, Phys. Rev. Lett. {\bf 119}, 235301 (2017).

\bibitem{prd80} Y.M. Cho, Phys. Rev. {\bf D21},
1080 (1980). See also Y. S. Duan and M. L. Ge,
Sci. Sinica {\bf 11},1072 (1979).
\bibitem{prl81} Y.M. Cho, Phys. Rev. Lett. {\bf 46},
302 (1981); Phys. Rev. {\bf D23}, 2415 (1981).

\bibitem{plb82} Y.M. Cho, Phys. Lett. {\bf B115}, 125 (1982).

\bibitem{prd00} Y.M. Cho, Phys. Rev. {\bf D62}, 074009 (2000).
\bibitem{jhep05} Y.M. Cho, M.L. Walker, and D.G. Pak,
JHEP {\bf 05}, 073 (2004); Y.M. Cho and M.L. Walker,
Mod. Phys. Lett {\bf A19}, 2707 (2004).
\bibitem{prd13} Y.M. Cho, Franklin H. Cho, and J.H. Yoon,
Phys. Rev. {\bf D87}, 085025 (2013); Y.M. Cho, Int. J.
Mod. Phys. {\bf A29}, 1450013 (2014).
\bibitem{epjc19} Y.M. Cho and Franklin H. Cho,
Euro. Phys. J. {\bf C79}, 498 (2019).

\bibitem{schw} J. Schwinger, Science {\bf 165}, 757 (1969).
\bibitem{bpst} A. Belavin, A. Polyakov, A. Schwartz,
and Y. Tyupkin, Phys. Lett. {\bf B59}, 85 (1975);
G. 't Hooft, Phys. Rev. Lett. {\bf 37}, 8 (1976).
\bibitem{plb79} Y.M. Cho, Phys. Lett. {\bf B81}, 25 (1979).

\bibitem{fadd} L. Faddeev and A. J. Niemi, Phys. Rev. Lett.
{\bf 82}, 1624 (1999); Phys. Lett. {\bf B449}, 214 (1999).
\bibitem{shab}S. Shabanov, Phys. Lett. {\bf B458},
322 (1999); {\bf B463}, 263 (1999); H. Gies, Phys.
Rev. {\bf D63}, 125023 (2001).
\bibitem{zucc} R. Zucchini, Int. J. Geom. Meth. Mod.
Phys. {\bf 1}, 813 (2004).
\bibitem{kondo} K. Kondo, S. Kato, A. Shibata,
and T. Shinohara, Phys. Rep. {\bf 579}, 1 (2015).

\bibitem{mpla16} Kyoungtae Kimm, J.H. Yoon, S.H. Oh,
and Y.M. Cho, Mod. Phys. Lett. {\bf A31}, 1650053 (2016).
\bibitem{ellis} J. Ellis, N.E. Mavromatos, and T. You,
Phys. Lett {\bf B756}, 29, (2016).
\bibitem{bpscm} F. Blaschke and P. Benes, Prog. Theor.
Exp. Phys. 073B03 (2018).

\end{references}
\end{document}